\documentclass[pra,preprint,showpacs]{revtex4}
\usepackage[centertags]{amsmath}
\usepackage[koi8-r]{inputenc}
\usepackage{amsfonts}
\usepackage{amssymb}
\usepackage{amsthm} 
\usepackage{newlfont}
\usepackage{epsfig}
\usepackage{amscd}
\usepackage{subcaption}
\usepackage{graphicx}
\usepackage{footnote}
\usepackage{lipsum}
\usepackage{color}
\usepackage{xcolor}
\usepackage{multirow}
\usepackage{amscd}

\newcommand{\beq}{\begin{equation}}
\newcommand{\eeq}{\end{equation}}
\newcommand{\ba}{\begin{array}}
\newcommand{\ea}{\end{array}}
\newcommand{\bea}{\begin{eqnarray}}
\newcommand{\eea}{\end{eqnarray}}
\begin{document}

\begin{center}
{\large \sc \bf {Nearest-neighbor approximation in one-excitation state evolution along
  spin-1/2 chain governed by $XX$-Hamiltonian
}}

\vskip 15pt

E.B.Fel'dman and A.I.Zenchuk


\vskip 8pt

{
Federal Research Center of Problems of Chemical Physics and Medicinal Chemistry RAS,
Chernogolovka, Moscow reg., 142432, Russia}.




\end{center}

\begin{abstract}

The approximation of nearest neighbor interaction (NNI)  is widely used in short-time spin dynamics with dipole-dipole interactions (DDI) when the  intensity of spin-spin  interaction  is  $\sim 1/r^3$, where $r$ is a distance between those spins. However,  NNI can not approximate the long time evolution in such systems. We consider the system with the intensity of the spin-spin interaction  $\sim 1/r^{\alpha}$, $\alpha\ge 3$, and find the low  boundary $\alpha_c$ of applicability of the NNI to the evolution of an arbitrary one-excitation initial quantum state  in the  homogeneous spin chain governed by the $XX$-Hamiltonian.  We obtain the logarithmic dependence of  $\alpha_c$  on the chain length.

\end{abstract}


\maketitle

\section{Introduction}
\label{Section:introduction}
{The coupling constants (exchange integrals) in systems with exchange interaction \cite{Y} decrease exponentially with the distance between the  interacting particles  \cite{LL}. Therefore only the interactions between the nearest neighbors are significant, while the exchange interaction between the remote particles can be neglected.}
{Both models, including nearest neighbor approximation (NNI) and all-neighbor interaction (ANI) were used in solving various problems of spin dynamics, including the dynamics of spin system with dipole-dipole interaction (DDI). Thus the well-known perfect state transfer was obtained for the properly engineered linear spin chain governed by the $XX$-Hamiltonian with NNI \cite{CDEL,KS}. However, the high probability state transfer was achieved for both NNI and ANI models \cite{GKMT,CRMF,ZASO,ZASO2,ZASO3,FZ_2009,DFZ_2009}. Similarly, both NNI and ANI where used in the analysis of quantum correlations, in particular, entanglement \cite{VGIZ,AFOV,HHHH}.  Although most of the above protocols can be used for any model of particle interaction, the result depends (sometimes significantly) on the adopted model. The interest to the NNI is provided, first of all, by the fact that it allows to significantly simplify numerical simulations allowing to apply the Jordan-Wigner transformation  \cite{JW} and yields exact solution in some cases.}

 A set of theoretical methods of spin dynamics in one-dimensional systems  with DDI  is based on the approximation of NNI  \cite{Abragam}. It can be used for interpretation of the  magnetic resonance experiments in spin chains (the NMR line-shape \cite{ALC,BFV}, multiple-quantum dynamics \cite{DMF,CRC}, quantum correlations  (entanglement and discord) \cite{KKWC,YLLF},  quantum state transfer \cite{Z_2014,FPZ_2021}). However, NNI is applicable only over  relatively short evolution time intervals when $Dt\leq 1$ ($D\sim 1/r^3$ is the coupling constant of dipole-dipole interaction between nearest neighbors). This  condition is not satisfied, for instance,  in state transfer along a long spin chain, when the remote-spin interactions become significant. We also mention results of Refs.\cite{FKZ_2010,FZ_2022} where some restrictions for  applicability 
of NNI were demonstrated. { In our recent paper \cite{FZ_2022} we compare the results  of the quantum state transfer along the zigzag and alternating spin chains governed by the XXZ Hamiltonian using both NNI and ANI. }   { In particular, it was shown  that the spin dynamics  governed by the XXZ-Hamiltonian with DDI can not be described by the NNI and requires including interactions with at least several neighboring spins. For instance, the end-to-end one-qubit  excited state transfer along the zigzag chain of 40 nodes requires including  at least 14-neighbor interactions, while similar transfer along the alternating chain requires using ANI {\cite{FZ_2022}}.}

{The problem of applicability of the NNI to DDI is important for the systems of arbitrary dimension and various geometry. As the first  step in studying this problem we consider the one-dimensional system (the open homogeneous spin chain) with $XX$-interaction \cite{JW}.}
We consider the case, when the above mentioned coupling constant obeys the rule  $D\sim 1/r^{\alpha}$ ($\alpha\ge 3$), which results in reducing the role of remote spin interactions. We study (i) the one-excitation end-to-end state transfer and (ii) evolution of an arbitrary one-excitation initial state  in the homogeneous chain of $N$ spins  ($s=1/2$) with XX-Hamiltonian \cite{JW,FR_2005,KF_2006}. We compare the probability amplitudes calculated using both  NNI and ANI over a chosen time interval using the special integral characteristics  and find the critical value $\alpha_c$ such that the NNI is applicable for $\alpha\ge \alpha_c$. The logarithmic dependence of $\alpha_c$ on the chain length $N$ is demonstrated.

\section{Hamiltonian and evolution}
\label{Section:Ham}
The $XX$-Hamiltonian taking into account the interaction of $M$ ($1\le M\le N-1$) nearest neighbors in the open spin chain  reads  \cite{FZ_2022}:
\begin{eqnarray}\label{XX}
H_M=\sum_{i=1}^{N-1} \sum_{j=i+1}^{\min(i+M,N)}D_{ij} (I_{xi}I_{xj} + I_{yi}I_{yj}), \;\;D_{ij}=\frac{\gamma^2 \hbar}{r_{ij}^{\alpha}},\;\;\alpha\ge 3,
\end{eqnarray}
where $\gamma$ is the gyromagnetic ratio, $\hbar$ is the Plank constant, $r_{ij}$ is the distance between the $i$th and $j$th spins and {$I_{\beta i}$ ($\beta=x,y$) are the $\beta$-projection operators of the $i$th spin momentum.}
{  We shall notice that the Hamiltonian (\ref{XX}) doesn't presume any particular requirement to the geometry of an open chain. In particular, the system might be 1-, 2- or 3-dimensional. However, if $M<N-1$, one has to make sure that the chosen $M$-neighbor  approximation is applicable. }
{Hereafter we work with the line homogeneous spin chain and}  set either  $M=1$ (NNI) or $M=N-1$ (ANI) {and consider the problem of applicability of NNI to approximate the real physical systems with  spin-interaction}. Below we use the dimensionless time $\tau =D_{12} t$. {We emphasize that the  proposed protocol can be applied to verify the applicability of the NNI to the dynamics { in the open spin chain}  governed by any Hamiltonian. Of course, the  result depends on the particular choice of the Hamiltonian.}

 Let us consider the dynamics of some quantity $F(\rho(\tau))$ under the  XX-Hamiltonian (\ref{XX})  over the fixed  $\tau$-interval $T$.  We use the subscript $;M$ to indicate that the quantity $F$ is found using the $M$-neighbor interaction. For instance,   $F_{;1}$  and  $F_{;N-1}$ mean that $F$ is calculated using, respectively,  the NNI and ANI. Obviously,   $F\equiv F_{;N-1}$. 
 { If we need to obtain the value of the function $F$ at the particular time instant $t_0$, then, to check whether the NNI is applicable for approximating the exact value $F(t_0)$, we can relay on the inequality 
 \begin{eqnarray}
 \frac{|F_{;N-1}(t_0) - F_{;1}(t_0)|}{F_{;N-1}(t_0)}<\varepsilon,
 \end{eqnarray}
 where $\varepsilon\ll 1$ is some conventional small parameter. However, this inequality doesn't give any information about relation between $F_{;N-1}$ and $F_{;1}$ at any other time instant $t<t_0$ and $t>t_0$.}
 To verify whether NNI can be used to  approximate the exact evolution of $F$  over the chosen time interval $T$, we follow Ref.\cite{FZ_2022} and introduce the following integral characteristics: 
\begin{eqnarray}
\label{JF2}
&&
\Delta J(F)=
\frac{J(F_{;N-1}-F_{;1})}{J(F_{;N-1})},\\
\label{JF}
&&J(F) =  \sqrt{\frac{1}{T}\int_{0}^T |F_{;N-1}(\tau)|^2 d\tau}.
\end{eqnarray}
We say that the NNI is applicable if 
\begin{eqnarray}\label{Fvar}
\Delta J(F)< \varepsilon,
\end{eqnarray}
where $\varepsilon \ll 1$ is the above mentioned conventional small parameter. We call the quantity 
$\Delta J$  the relative  discrepancy between the NNI and ANI protocols. By definition, it depends on the  considered $\tau$-interval $T$. 

If the set of $K$ quantities $\mathbf{F} =\{F^{(n)}: n=1,\dots, K\}$ must be studied over the same $\tau$-interval $T$, then  the maximum of $\Delta J(\mathbf{F})$ must satisfy condition (\ref{Fvar}), i.e.,
\begin{eqnarray}\label{JFF}\label{Jvar}
\Delta J(\mathbf{F}) \equiv \max_{n=1,\dots,K} \Delta J(F^{(n)}) <\varepsilon.
\end{eqnarray}

\section{Evolution of one-excitation state}
\label{Section:ev}
We  consider the evolution of an arbitrary  one-excitation initial state $|\psi(0)\rangle$ over the time interval $T$, thus
\begin{eqnarray}
\label{init}
|\psi(0)\rangle = \sum_{j=1}^N a_{j}|j\rangle,\;\;\sum_{j=1}^N |a_{j}|^2=1 .
\end{eqnarray}
Then
 we have 
\begin{eqnarray}
\label{psi2}
&&
|\psi(t)\rangle_M = e^{-i \frac{H_M}{D_{12}} \tau} |\psi(0)\rangle = 
\sum_{j,k=1}^N  a_{j} p_{jk;M}(t) |k\rangle,\;\;\tau=D_{12} t\\\nonumber
&&p_{jk;M}= \langle k|e^{-i \frac{H_M}{D_{12}} \tau}|j\rangle,\;\;
\sum_{k=1}^N p_{ik;M} p^*_{jk;M} =\delta_{ij},
 \end{eqnarray}
where $|j\rangle$ means the state with the single $j$th excited spin, { $p_{jk;M}$ means the probability amplitude $p_{jk}$ calculated using the Hamiltonian $H_M$, star means complex conjugate and $\delta_{ij}$ is the Kronecker symbol.}
If, in particular, we are interested in the end-to-end one-excitation state transfer, then the initial state (\ref{init}) and evolution (\ref{psi2}) reduce, respectively, to 
\begin{eqnarray}
\label{init1}
|\psi(0)\rangle =|1\rangle 
\end{eqnarray}
and
\begin{eqnarray}
\label{psi3}
&&
|\psi(t)\rangle_M = 
\sum_{k=1}^N  p_{1k;M}(t) |k\rangle,\;\;
\sum_{k=1}^N p_{1k;M}=1.
 \end{eqnarray}


Now we consider  $\alpha\ge 3$ in (\ref{XX})  and study the NNI in application to  the evolution of two functions. First one is a scalar function 
\begin{eqnarray}\label{F}
&&
 F=p_{1N}\;\;{\mbox{with}} \;\;T=2 N.
\end{eqnarray}
It is associated with the end-to-end one-excitation state transfer, Eqs.(\ref{init1}), (\ref{psi3}).  The choice $T=2N$ in (\ref{F}) is rather conventional, it is about twice the  time interval of excited state transfer {(along a homogeneous spin chain)} which is  $\sim N$ {for $XX$-Hamiltonian}. 
The second function is prompted by the process of evolution of an arbitrary initial state given by 
Eqs.(\ref{init}), (\ref{psi2}).  According to Eq.(\ref{psi2}), NNI is applicable to the  considered evolution if it is applicable to the evolution of each probability amplitude $p_{jk}$. Therefore we can introduce the vector function
\begin{eqnarray}
\label{bfF}
&&
 \mathbf{F}=\{p_{ij}: i,j=1,\dots, N\}\;\;{\mbox{ with}}\;\; T=4N.
\end{eqnarray}
The choice  $T=4N$ in (\ref{bfF}) is also conventional, it is about  twice the returning time of the excited state of the 1st spin. 

{We proceed with the simplest case of the end-to-end state transfer associated with  the function $F$ (\ref{F}).}
Substituting  $F=p_{1N}$  and $T=2N$  into Eq.(\ref{JF2}) and (\ref{JF}) we obtain, taking into account  condition (\ref{Fvar}),
\begin{eqnarray}\label{J1N}
\Delta J(p_{1N})&=&
\sqrt{\frac{\int_{0}^{2N} |p_{1N;N-1}(\tau)-p_{1N;1}(\tau)|^2d\tau}
{\int_{0}^{2N} |p_{1N;N-1}(\tau)|^2 d\tau}} <\varepsilon.
\end{eqnarray}

\begin{figure*}[!]
\begin{subfigure}[h]{0.4\textwidth}
\includegraphics[scale=0.45]{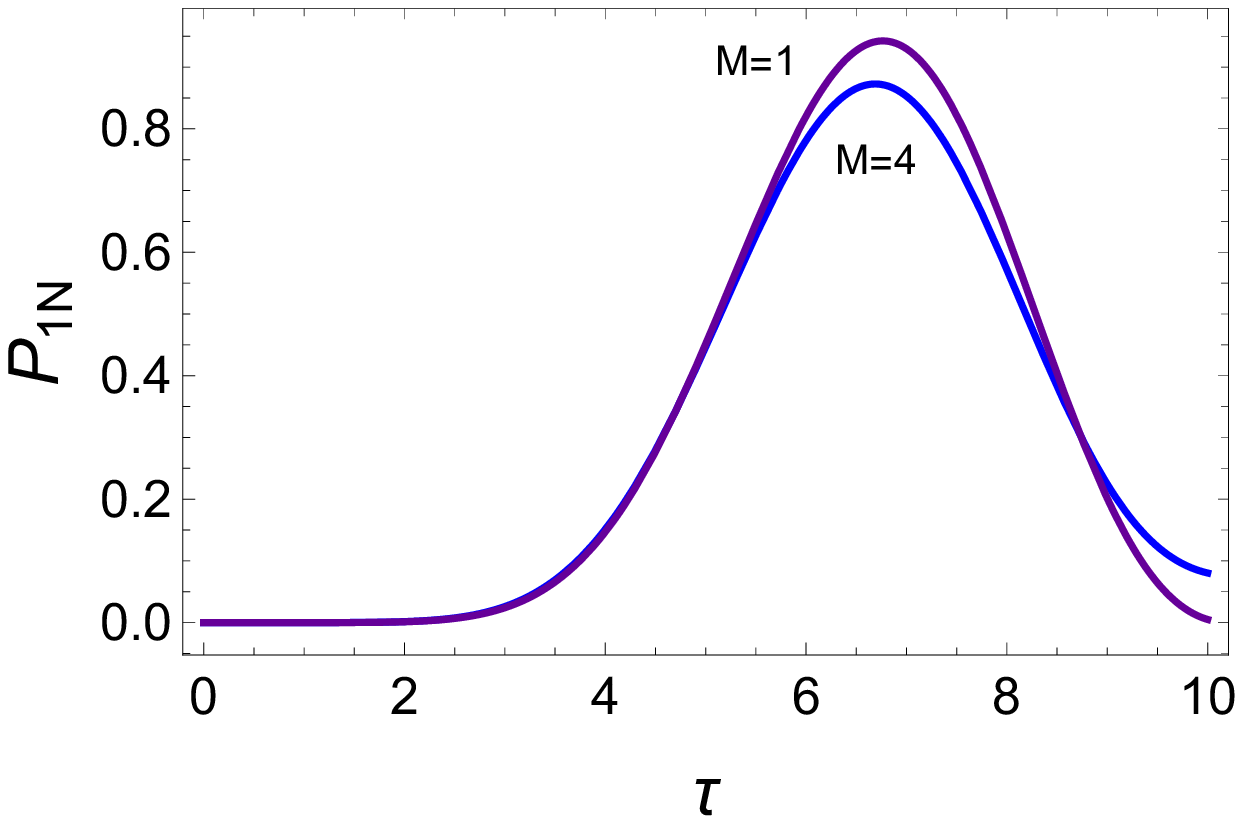}
 \caption{$N=5$}
\end{subfigure}
\begin{subfigure}[h]{0.4\linewidth}
\includegraphics[scale=0.45]{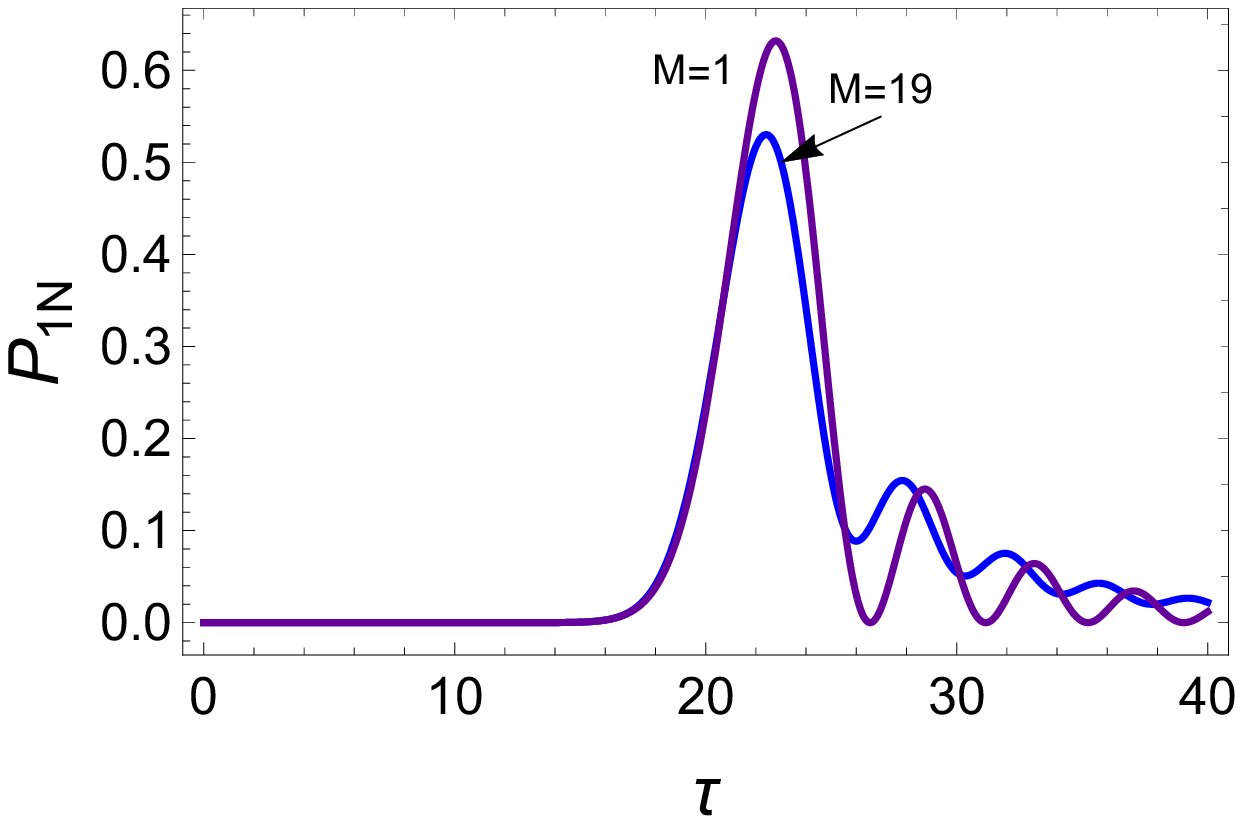}
 \caption{$N=20$}
\end{subfigure}
\caption{The state transfer probability $P_{1N;M}$ calculated using ANI, $M=N-1$, and NNI, $M=1$, {for the evolution governed by Hamiltonian (\ref{XX}) with DDI ($\alpha=3$)}. (a) $N=5$ and (b) $N=20$. }
\label{Fig:NNI}
\end{figure*}

\begin{figure}
\includegraphics[scale=0.5]{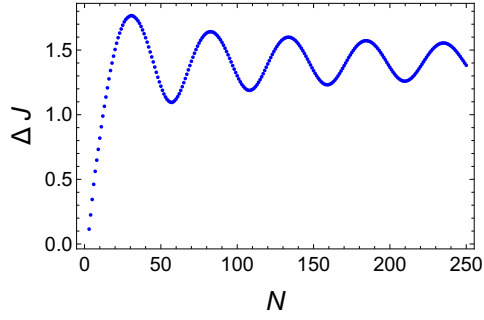}
\caption{The discrepancy $\Delta J(p_{1N})$ for {the Hamiltonian (\ref{XX})  with} $\alpha=3$ as a function of the chain length $N$.}
\label{Fig:J}
\end{figure}

As was mentioned in the introduction, the NNI can approximate a  short-time dynamics 
of the   spin system with DDI ($\alpha=3$) governed by the XX-Hamiltonian. As  a justification of this fact, we show the graphs of the probabilities of the end-to-end excited state transfer $P_{1N;M}=|p_{1N;M}(t)|^2$ along a relatively  short spin chains ($N=5$ and 20)  in Fig.\ref{Fig:NNI}  for both NNI and  ANI (i.e., $M=1$ and $M=N-1$) over the time interval $T= 2 N$, { $\alpha=3$}.
Comparing discrepancies between two curves in Fig.\ref{Fig:NNI}a and Fig.\ref{Fig:NNI}b we see that the approximation by NNI becomes worse for the longer chain.
This is confirmed by the graph of $\Delta J(p_{1N})$ as a function of $N$ given in Fig.\ref{Fig:J} where $T=2N$. It demonstrates the significant discrepancy between the state-transfer probabilities obtained via  NNI and ANI models for a long chain.

We shall emphasize that Eq.(\ref{J1N})  characterizes not just  probability of the excited state registration at the $N$th spin at the optimal time instant, but the deviation $p_{1N;N-1}$ from $p_{1N;1}$ over the whole $\tau$-interval $T=2N$.

\begin{figure}
\includegraphics[scale=0.5]{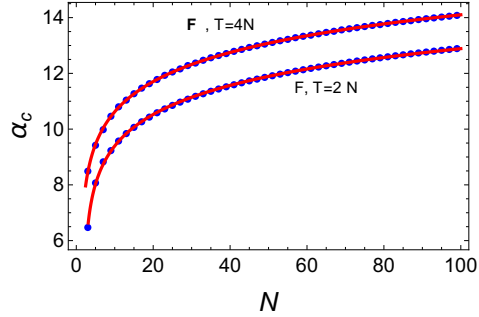}
\caption{The parameter $\alpha_c$  as a function of $N$ for the function $F$ from (\ref{F}) and $\mathbf{F}$ from  (\ref{bfF}) approximated by logarithm (\ref{alplog}) with the parameters, respectively,   (\ref{alplogpar}) (for $F$) and  (\ref{alp2log}) for $\mathbf{F}$.}
\label{Fig:alpc}
\end{figure}

{Now we proceed to study the applicability of the NNI to the more general case of evolution of an arbitrary initial state.} Substituting   $\mathbf{F}$  from (\ref{bfF}) into Eq.(\ref{Jvar}) requires calculating $\Delta J(p_{jk})$ using (\ref{JF2}), (\ref{JF}) with $T=4N$ which yields
\begin{eqnarray}
\Delta J(p_{jk}) = 
\sqrt{\frac{\int_{0}^{4N} |( p_{jk;N-1}(\tau)-p_{jk;1}(\tau)) |^2d\tau}
{\int_{0}^{4N} |p_{jk;N-1}(\tau)|^2 d\tau}} .
\end{eqnarray}
Then Eq.(\ref{Jvar}) yields 
\begin{eqnarray}\label{J1N2}
\Delta J(\mathbf{F})=\max_{ j,k=1,\dots,N} { \Delta J(p_{jk})} <\varepsilon.
\end{eqnarray}
We  say that the NNI is applicable for describing the evolution of $F$ from (\ref{F}) or $\mathbf{F}$ from (\ref{bfF})  if 
condition (\ref{J1N}) or (\ref{J1N2}) is satisfied.
Numerical simulations prompt us to choose $\varepsilon=0.01$. 
 Therefore, we may find such critical $\alpha_c$, that   condition (\ref{J1N}) (or (\ref{J1N2})) is satisfied for any $\alpha\ge \alpha_c$.  The graphs of  $\alpha_c$ as functions of $N$ are shown in Fig.\ref{Fig:alpc} for the both  above cases. {By definition, $\alpha_c$ for $F$ can not be bigger than $\alpha_c$ for $\mathbf{F}$ because $p_{1N}$ is in the list of probability amplitudes defining $\mathbf{F}$, see Eq.(\ref{bfF}). This fact is reflected in Fig.\ref{Fig:alpc}.}

\begin{figure}
\includegraphics[scale=0.5]{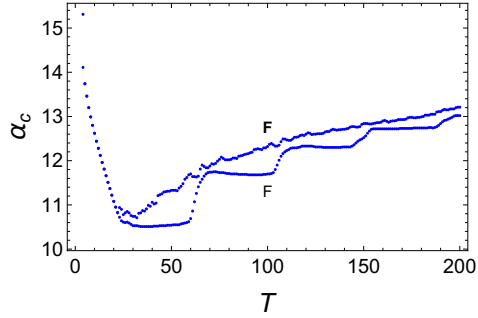}
\caption{The parameter $\alpha_c$  as a function of $T$ for the chain of $N=20$  spins and functions $F$ from (\ref{F}) and $\mathbf{F}$ from (\ref{bfF}).}
\label{Fig:T1}
\end{figure}

It is interesting that { each of two sets} of points $\alpha_c(N)$ {associated with the functions $F$ and $\mathbf{F}$} can be well approximated by the logarithm
\begin{eqnarray}\label{alplog}
\alpha_c(N)\approx a  \ln(N-b)+ c,
\end{eqnarray} 
using the least-squares method.
Here
\begin{eqnarray}\label{alplogpar}
&&a=1.393,\;\;b=2.005,\;\;c= 6.500\;\;{\mbox{for $F$ from}} \;\; (\ref{F}),\\
\label{alp2log}
&&
a = 1.434,\;\;  b = 1.496, \;\; c = 7.841\;\;{\mbox{for $\mathbf{F}$ from}} \;\;  (\ref{bfF}).
\end{eqnarray}
Obviously, an increase in $T$ leads to an increase in $\alpha_c$. For instance, $\alpha_c(T)$ at $N=20$ is shown in Fig.\ref{Fig:T1} for $F$ from (\ref{F})  and $\mathbf{F}$ from (\ref{bfF}).
{We observe steps in $\alpha_c$ for the case of $T$ which corresponds to the time instants of excited state returning to the $N$th node, these time instants $\sim (2 n+1) N$, $n=0,1,2,\dots$.}  On the contrary, the behavior  of $\alpha_c$ for  $\mathbf{F}$ is smooth. Two curves tend to each other with an increase in $T$.
\begin{figure}
\includegraphics[scale=0.5]{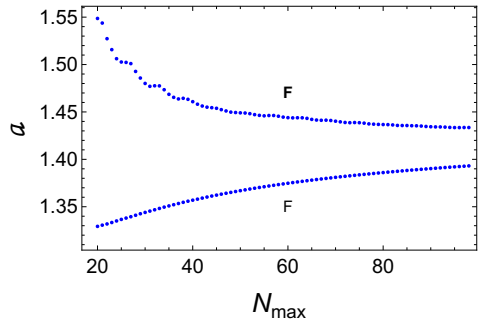}
\caption{The parameter $a$ in (\ref{alplog}) as a function of $N_{max}$ for the  functions $F$ from (\ref{F}) and $\mathbf{F}$ from (\ref{bfF}).}
\label{Fig:a}
\end{figure}

We shall give some remarks on  logarithm (\ref{alplog}). All the parameters $a$, $b$ and $c$ in (\ref{alplog})  depend on the maximal $N$ (or $N_{max}$) in the considered range of the chain lengths $N\le N_{max}$. For instance, $N_{max}=100$ in Fig.\ref{Fig:alpc}. However, the  parameter $a$ is the  most important one because it determines the principal growth of the logarithm with $N$. The graphs of $a(N_{max})$ are given in Fig. \ref{Fig:a} for both cases $F$ from (\ref{F}) and $\mathbf{F}$ from (\ref{bfF}). These graphs show principal difference between two considered cases: $a(N_{max})$ corresponding to $F$ is an increasing function, while $a(N_{max})$ corresponding to $\mathbf{F}$ is a decreasing  function. {Since $\alpha_c(\mathbf{F})\ge \alpha_c(F)$, these two curves can not cross each other and tend to some asymptotic value as $N_{max}\to\infty$.}

We shall notice that  the state-transfer amplitudes $p_{jk}$
yielding maximum in (\ref{J1N2}) correspond to the particular transitions between two nodes, which { have been found numerically. These transitions are marked by  bars in Fig.\ref{Fig:tr}}. We see that all such transitions are near the boundaries of a spin chain, {holding the symmetry $i\leftrightarrow  N-i+1$. Dots in the left-low corner (short chains) of this figure mean that the source node coincides with the target one.
\begin{figure}
\includegraphics[scale=0.7]{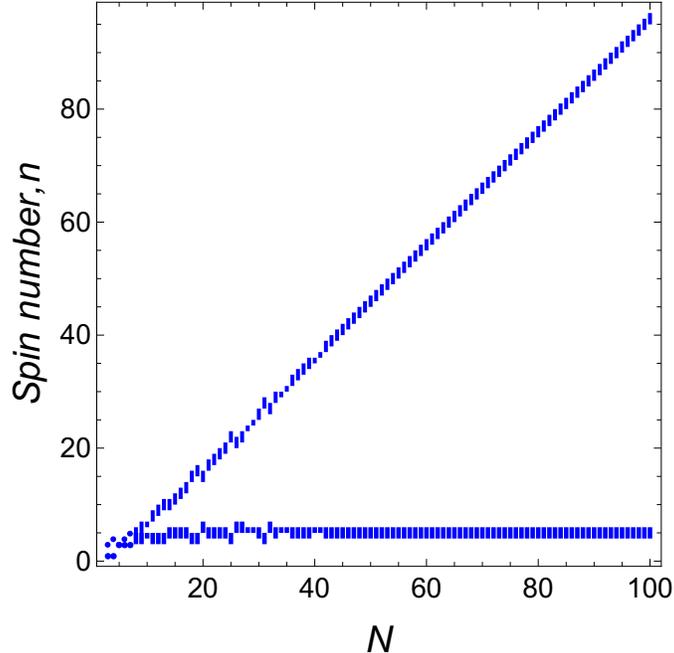}
\caption{Bars connect the $i$th and $j$th spins such that the transition amplitudes  $p_{ij}$   correspond  to the maximum in Eq.(\ref{J1N2}).
}
\label{Fig:tr}
\end{figure}

{
\subsection{Applicability of the proposed protocol to other models}

We shall emphasize, that the integral characteristics (\ref{JF2}), (\ref{JF}) can be used for analysis of applicability of the NNI to any particle configuration governed by any Hamiltonian because, introducing that characteristics, we do not use any particular property of the $XX$-Hamiltonian (\ref{XX}). Therefore, replacing the Hamiltonian, we can obtain results similar to those demonstrated in Figs. \ref{Fig:NNI}-\ref{Fig:a} of Sec.\ref{Section:ev} up to the quantitative discrepancies. The logarithmic dependence of $a_c$ on $N$ given in (\ref{alplog})  is obtained for the $XX$-Hamiltonian (\ref{XX}) and, at the moment, we don't have justifications that it holds for other models. }

{\section{Conclusion}}

{ We demonstrate that, in general, the NNI and ANI  are quite different models. Only in certain  cases of fast decaying coupling constants these two models yield equivalent results. That was observed for the $XX$-interaction with the coupling constants $\sim 1/r^\alpha$, $\alpha\ge \alpha_c$.
}

Thus, although  NNI in application to the spin evolution with DDI governed by the XX-Hamiltonian is a reasonable approximation over short time interval, it yields a significant error in general.
Nevertheless, the NNI is applicable for describing certain processes in spin dynamics over desired time interval $T$ if the parameter $\alpha$ in (\ref{XX}) is large enough.  We find the critical value (lower boundary) $\alpha_c$ as a function of the spin length $N$ for the process of the end-to-end excited one-qubit pure state transfer over the time interval $2N$ and for the spin dynamics with an arbitrary initial state over the time interval $4N$. In both cases $\alpha_c(N)$
can be approximated by the logarithmic function, and $\alpha_c(N) \to a \ln (N)$ as $N\to \infty$.
The form of the function $a(N_{max})$ depends on a particular function under consideration (scalar $F$ or vector $\mathbf{F}$ in our case, see Fig.\ref{Fig:a}).

{\bf Acknowledgments} The work was performed as a part of a state task, State Registration No. AAAA-A19-119071190017-7.

\end{document}